\documentclass[11pt,oneside,a4paper]{article}
\usepackage{graphicx,caption,subcaption,array,authblk}
\usepackage{amsmath}
\usepackage[parfill]{parskip}
\usepackage[margin=1.0in]{geometry}
\usepackage[backend=bibtex,style=authoryear]{biblatex}
\usepackage[outdir=./]{epstopdf}
\usepackage{lineno,setspace}

\usepackage{hyperref}
\usepackage[noabbrev,nameinlink]{cleveref}
\creflabelformat{equation}{#2\textup{#1}#3}

\makeatletter
\newcommand{\thickhline}{%
    \noalign {\ifnum 0=`}\fi \hrule height 1pt
    \futurelet \reserved@a \@xhline
}
\makeatother
 
\title{ESCAPE – Echo SCraper and ClAssifier of PErsons: A novel tool to facilitate using voice-controlled devices for research}
    \date{\today}
    \author[1]{Nicholas C. Firth \thanks{nicholas.firth@ucl.ac.uk}} %
    \author[2]{Emma Harding}
    \author[3]{Mary Pat Sullivan}
    \author[2]{Sebastian J. Crutch}
    \author[1]{Daniel C. Alexander}
    \affil[1]{Centre for Medical Image Computing, Department of Computer Science, UCL}
    \affil[2]{Dementia Research Centre, University College London}
    \affil[3]{School of Human and Social Development, Nipissing University}

\begin{document}
    \maketitle

    \begin{abstract}
        Smart devices have become common place in many homes, and these devices can be utilized to provide support for people with mental or physical deficits. Voice-controlled assistants are a class of smart device that collect a large amount of data in the home. In this work we present Echo SCraper and ClAssifier of Persons (ESCAPE), an open source software for the extraction of Amazon Echo interaction data, and speaker recognition on that data. We show that ESCAPE is able to extract data from a voice-controlled assistant and classify with accuracy who is talking, based on a small number of labeled audio data. Using ESCAPE to extract interactions recorded over 3 months in the first author's home yields a rich dataset of transcribed audio recordings. Our results demonstrate that using this software the Amazon Echo can be used to study participants in a naturalistic setting with minimal intrusion. We also discuss the potential for usage of voice-controlled devices together with ESCAPE to understand how diseases affect individuals, and how these data can be used to monitor disease processes in general.
    \end{abstract}

    \section{Introduction} 
    \label{sec:introduction}
    Over the past five years the Internet of Things (IoT), the networked interconnection of everyday objects, which are often equipped with ubiquitous intelligence \parencite{xia_internet_2012}, has increased in popularity. More recently, major advances in voice recognition have led to commercial voice assistive technology such as the Google Home and Amazon Echo. The Amazon Echo is currently the most popular voice-controlled device in the home, with 5.1 million devices sold by November 2016 \parencite{levin_amazon_2016}. The Echo and Echo Dot are smart speakers which connect to the intelligent assistant system Alexa. The Echo waits until the wake word (Alexa by default) is spoken and then listens for a command. The device is capable of making shopping and to-do lists, music playback, setting alarms and timers, and providing real-time information such as weather and news. All interactions, including audio data, with an Echo are saved by Amazon to improve experience and services, though this can be deleted by individuals who do not wish for their data to be retained.

    Using new commercial devices for research poses practical challenges such as effectively mining the data and analyzing it in a meaningful way. For an Echo device one of the main challenges is extracting the data from the Amazon servers in a systematic manner that does not violate terms of use agreements with Amazon. Another key challenge is to identify who is speaking in a given saved interaction. Whilst it is possible to manually code who is speaking in a small number of audio recordings, this is unfeasible with a large number of recordings due to the time it takes to generate accurate labels. One method used to annotate large amounts of data is crowdsourcing. Crowdsourcing reduces the time taken by an individual, but is not a free service and it has been shown to introduce errors into data labels \parencite{albarqouni_aggnet:_2016}.

    One potential application of voice assistive technology is in the healthcare sector. Current research in IoT in healthcare is focused on developing assistive technology, such as fall detection in care homes \parencite{greene_iot-based_2016}, and wandering detection in people with Alzheimer's disease \parencite{ng_not_2016}. Integration of such devices into locations such as care homes will generate a large amount of data describing the actions of people in their day-to-day life rather than in constrained experimental conditions. Such data may provide valuable insight into diseases such as Alzheimer's and Multiple Sclerosis to better quantify how such diseases affect people in day-to-day life. Transcribed audio data collected in the home has the potential to detect subtle changes in memory and verbal skills \parencite{rankine_language_2017}. These subtle changes are relevant to people living with conditions such as Multiple Sclerosis, Alzheimer's disease, and aphasic symptoms due to a stroke.

    In the U.K., there was a 55\% increase in the number of people diagnosed with dementia between 2010 and 2016 \parencite{aruk_diagnoses_????}. With the annual cost of care rising from \pounds26,000 for mild dementia to \pounds55,000 for severe dementia, there is a massive financial and societal benefit for people to ‘live well’ with dementia for as long as possible. There is, as yet, no disease modifying therapeutic intervention for most dementia pathologies, and ability to perform activities of daily living is the main factor affecting quality of life in patients with dementia \parencite{andersen_ability_2004}. While there are some existing technology-based interventions, these are rarely designed for people with dementia, in terms of the cognitive capacity required for the whole process of operation. As such they often require assistance to use/set up.

    Using the Amazon Echo to collect transcribed audio data provides an inexpensive method of longitudinal data collection within a naturalistic setting, which minimizes both intrusiveness and demand on those living with dementia. As well as a tool for collecting data, devices such as the Echo have the potential to positively impact upon the lives of both someone living with dementia and caregivers. For people living with memory-led Alzheimer’s disease could direct repetitive questions towards the Amazon Echo and potentially reduce frustration for themselves and/or family members, and be reminded of important events, such as taking one's medication. For people living with the progressive visual dementia posterior cortical atrophy (PCA) \parencite{crutch_consensus_2017} the Echo could remove potentially challenging interfaces such as television remotes and mobile phones.

    In this paper, we will describe the Echo SCraper and ClAssifier of Persons (ESCAPE) software which has been developed to quickly and systematically download data collected by an Amazon Echo and then perform speaker recognition on each interaction. We hope that ESCAPE will enable further research, in the areas of dementia, other healthcare contexts and beyond, using the Amazon Echo, which we argue is a valuable resource both to neurological and wider research communities.

    \section{Methods} 
    \label{sec:methods}

    \subsection{Data Mining} 
    \label{sub:data_mining}
    Interactions with an Amazon Echo are recorded and stored by Amazon and these recordings can be accessed by users online, or on their Alexa phone/tablet application, however these interaction recordings are not available for bulk download. To systematically retrieve all interactions, it was necessary to produce an automated service to download interactions from the Alexa system. Software that automatically mines data from web pages is termed web scraping. Web scraping typically downloads web sites, extracts information and uses this information to download subsequent websites to extract more information \parencite{mitchell_web_2015}. In this work a web scraper was written to access and download all data from an Amazon account (which can be associated with many Echoes). The scraper requires an HTTP cookie copied from a web browser after signing in to an Amazon account on the Alexa website. Once this has been copied the web scraper will download all audio recordings, transcripts and other details required by the Echo to function, for the account that is signed in. 

    \subsection{Speaker Recognition}
    To effectively use the Echo to study individuals it is necessary to perform speaker recognition on audio recordings. Two methods were used for speaker recognition: Kullback-Leibler distances were used to identify a diverse set of audio recordings to annotate, and Hidden Markov Models were subsequently used to classify labeled and unlabeled data. For both of these methods audio files longer than 1.5 seconds were truncated to this length, as this is typically when the user says the wake word, Alexa. All audio was then decomposed into their Mel-Frequency Cepstral Coefficients (MFCCs).

    \subsubsection{Mel-Frequency Cepstral Coefficients} 
    \label{ssub:mel_frequency_cepstral_coefficients}
    Mel-frequency cepstrum are a representation of the power spectrum on the Mel-scale, which is a subjective scale for the measurement of pitch constructed from determinations of the half-value of pitches at various frequencies \parencite{volkmann_scale_1937}. Mel-frequency cepstral coefficients (MFCCs) are the coefficients that make up the Mel-frequency cepstrum. To compute the MFCCSs from audio, first a window function was applied to audio data to quantize the data into small overlapping frames. A one dimensional discrete Fourier transform was then applied to each of the frames independently. 26 frequencies equally spaced according the Mel-scale, between 0Hz and the Nyquist frequency (8 kHz for audio recorded from an Echo) were then defined. A weighted sum of the Fourier transform magnitudes around each of these frequencies was computed, then the natural log of each of the 26 sequences was taken. A discrete cosine transform was then applied to remove correlation, and the first 13 coefficients were kept for each time step. This results in a time series of length $L$,

    \begin{align*}
     L = (TS / W_s) - (W_l/W_s) -1
    \end{align*}
    where $TS$ is the time in seconds (number of samples in recording divided by the sampling rate), $W_s$ is the window step and $W_l$ is the window length. Each sample in the time series contains the 13 MFCCs coefficients for that time step. In the current work MFCCs were generated using the python speech features library \parencite{_jameslyons/python_speech_features_????}, with default parameters (window length of 25 milliseconds, a window step of 10 milliseconds, 26 Mel-frequency bands and 13 cepstrum).

    \subsubsection{Kullback-Leibler Divergence} 
    \label{ssub:kullback_leibler_divergence}
    The Kullback-Leibler (KL) divergence is a measure of difference between two probability distributions. The KL divergence for two $k$-dimensional Gaussians, $\mathcal{N}_0$ and $\mathcal{N}_1$ is given by
    \begin{align*}
        D_\text{KL}(\mathcal{N}_0 \| \mathcal{N}_1) & = \frac{1}{2} \left( \mathrm{tr} \left( \Sigma_1^{-1} \Sigma_0 \right) + \left( \mu_1 - \mu_0\right)^\top \Sigma_1^{-1} ( \mu_1 - \mu_0 ) - k + \ln \left( \frac{\det \Sigma_1}{\det \Sigma_0}   \right)  \right). \\ \intertext{Where $\mu_i$ and $\Sigma_i$ are the vector mean and covariance of $\mathcal{N}_i$ respectively. As the KL divergence is non-symmetric in this work we used $D_{\text{KL}_\text{Sym}}$ where}
        D_{\text{KL}_\text{Sym}}(\mathcal{N}_0 \| \mathcal{N}_1) &= D_\text{KL}(\mathcal{N}_0 \| \mathcal{N}_1) + D_\text{KL}(\mathcal{N}_1 \| \mathcal{N}_0).
    \end{align*}
    
    \subsubsection{Classification Using Kullback-Leibler} 
    \label{ssub:classification_using_kullback_leibler}
    To identify and label a diverse set of audio recordings from both participants KL divergence was used. Each audio file was transformed into its MFCCs, then a multivariate (13 dimensional) Gaussian was fit on these, if the minimum KL divergence of a file was less than 50 of a manually classified file then it was automatically classified with the same label. The cutoff of 50 was selected to minimize the number of false positives. If the KL divergence was greater than 50 then the audio file was played and manually labeled. 

    \subsubsection{Hidden Markov Models} 
    \label{ssub:hidden_markov_models}
    A Hidden Markov Model \parencite{baum_statistical_1966} is a statistical model which assumes that the data being modelled is a Markov process with latent states. An HMM assumes that every emission (in this work, MFCCs from a given frame) belongs to a latent state, modelled as a probability distribution and that every state has a probability of transitioning to another state. The transition path is a Markov chain, so depends only on the current emission. Due to their ability to recover a sequence that is not directly observable HMM’s are frequently used in speech and speaker recognition, to model transitions in speech \parencite{singh_speech_2012}.

    To use HMMs to generate features for machine learning we inferred a separate model for each audio file. We then computed the log likelihood of every sequence under every other sequence’s HMM model, which resulted in a square similarity matrix. In this work all HMM modelling was performed using the hmmlearn package \parencite{_hmmlearn/hmmlearn_????}. Five state Gaussian HMM’s were inferred with the Viterbi algorithm \parencite{viterbi_error_1967}.

    \subsubsection{Machine Learning} 
    \label{ssub:machine_learning}
    The similarity matrix was standardized by removing the mean and scaling to unit variance. Principal components analysis wereperformed performed on the similarity matrix to visualize the data in a lower dimension. Principal component analyses was performed using the scikit-learn package \parencite{pedregosa_scikit-learn:_2011} in python with default parameters. A Ridge Regression Classifier was trained using nested cross-validation to simultaneously estimate the optimum regularization parameter and estimate the performance of this algorithm on unseen data. A Ridge Regression Classifier was chosen as the model is linear and enforces sparsity, thus prevents overfitting. 

    \section{Results}

    \subsection{Data} 
    \label{sub:data}
\begin{figure}[htb]
        \centering
        \begin{subfigure}[b]{0.575\textwidth}
            \includegraphics[width=\linewidth]{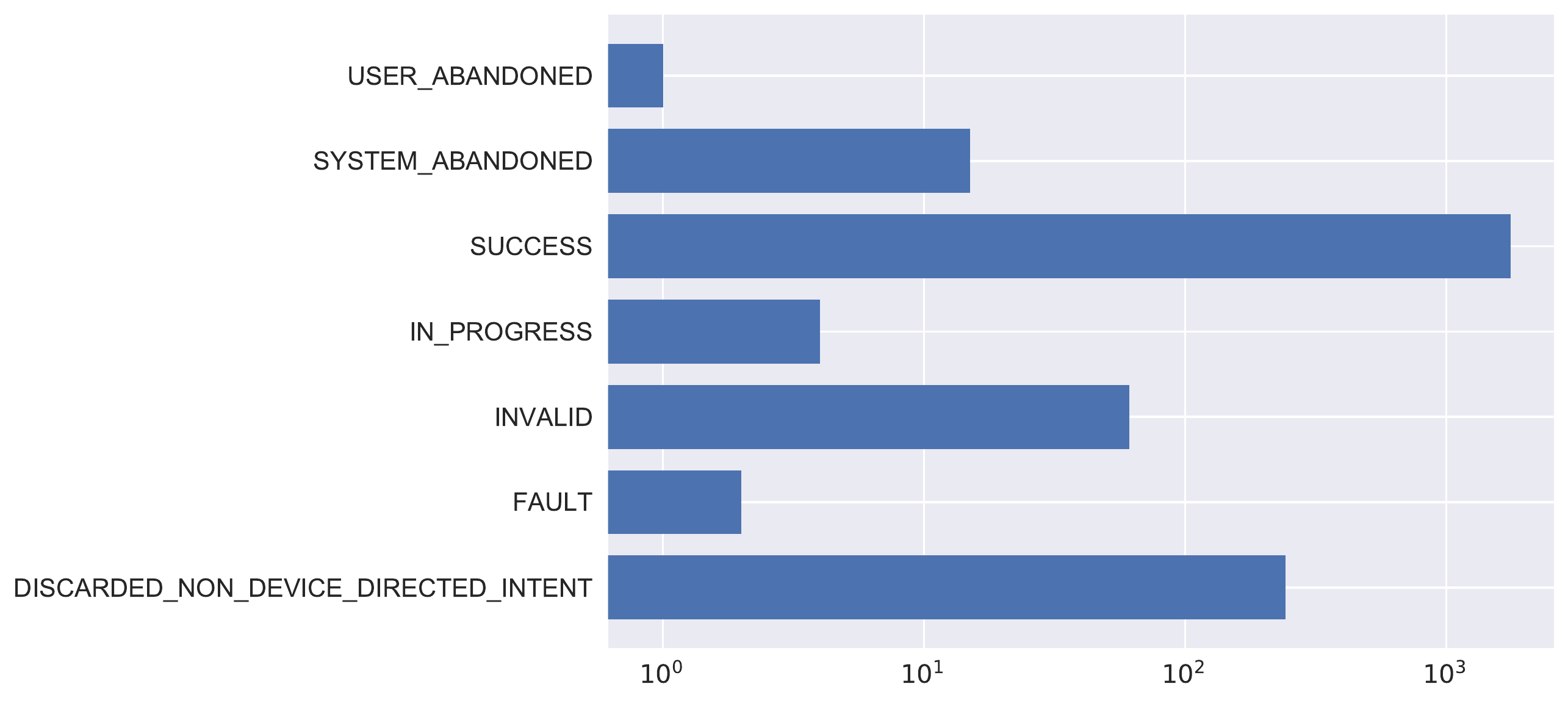}
            \caption{Counts of Amazon activity statuses for all interactions downloaded.}
            \label{fig:activity_stats}
        \end{subfigure}
        \begin{subfigure}[b]{0.41\textwidth}
            \includegraphics[width=\linewidth]{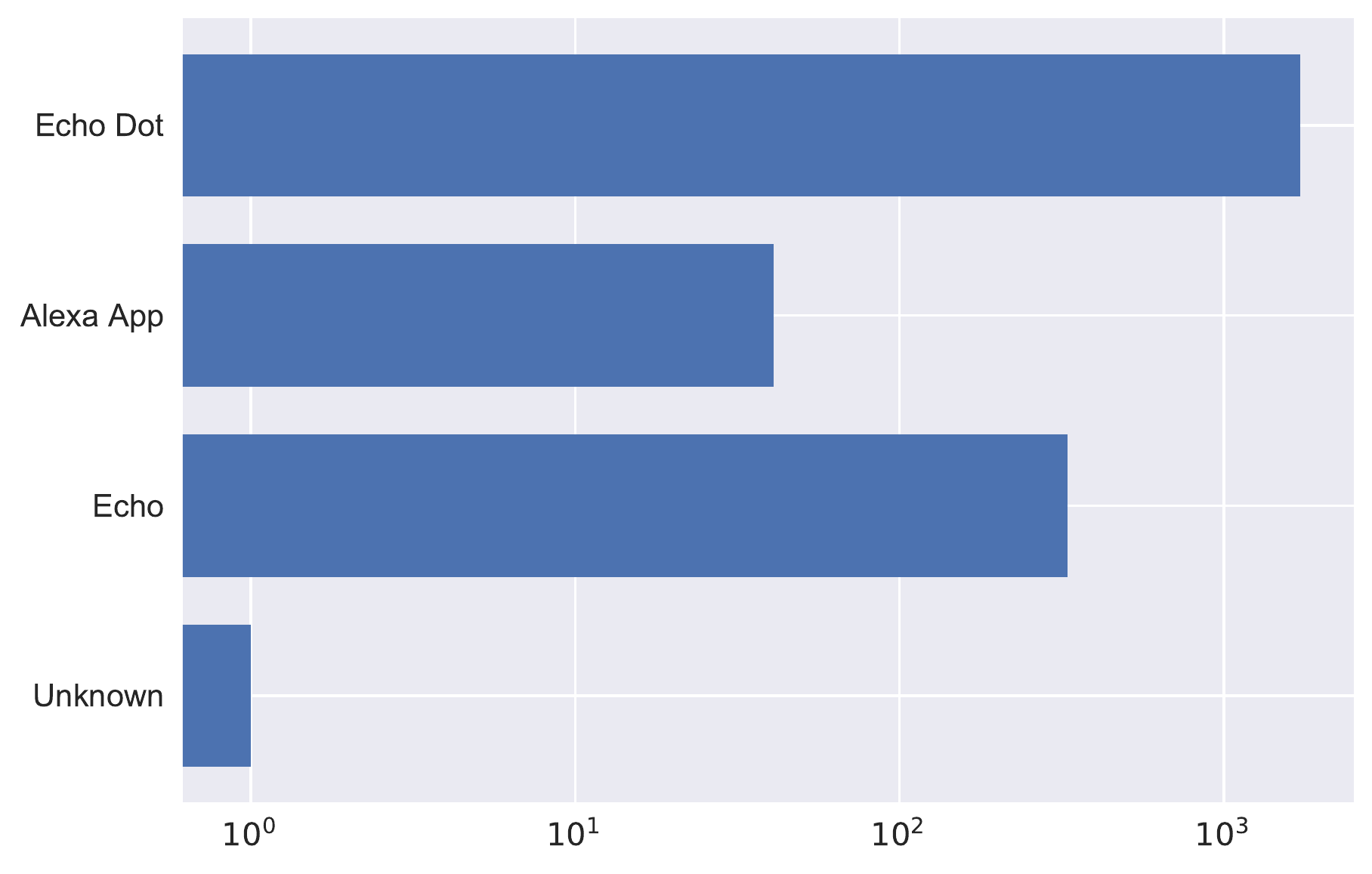}
            \caption{Counts of different devices used to interact with the Alexa system.}
            \label{fig:device_usage}
        \end{subfigure}
        \caption{Device usage counts.}

    \end{figure}

    One Amazon Echo Dot was installed for 106 days, and one Amazon Echo was installed for 35 days. These devices were installed in a kitchen (Echo Dot) and living room (Echo) with one male and one female residing in the home. All interaction data, including audio was collected for this period. In total 2,902 interactions were downloaded, of these 1,807 had both audio data and transcribed text, 232 had audio and no text and 5 had neither. Interaction data mined from the Alexa site shows that 1,767 (84.5\%) of attempted interactions with an Echo resulted in a ‘SUCCESS’ status (\Cref{fig:activity_stats}), i.e. the Alexa system performed an action based on what it understood was said. Analysis of the serial numbers used in the downloaded data shows that the majority of interactions were made with the Echo Dot (\Cref{fig:device_usage}), though the Echo Dot was installed for longer the average number of uses per day (16.2) is still higher than the Echo (9.4).
    
    \subsubsection{Experiments} 
    \label{ssub:experiments} 
    All interaction data and audio data was collected from the Amazon account associated with the two Echoes. To curate a rich dataset and demonstrate the potential use of the Echo for research, all transcribed audio recordings were labelled as Male (the first author), or Female. A machine learning approach was used to accurately label the data points whilst minimizing researcher’s time. First 185 audio recordings were manually labelled using the Kullback-Leibler labelling technique described above. The labelled audio files were then used to generate features for machine learning, using the MFCC and HMM methods described. Data was then split into stratified training and test sets (67:33), and all similarities to instances in the test set were removed from the feature set. This step was taken to ensure that no information relating to the test set was available to the model during training. Three-fold cross-validation was then used on the training set to select a regularization strength for a ridge regression classifier model, 20 strengths were sampled from $10^{-10}$ to $10^{10}$ in increasing powers of 10. The model was then refit on the entire training set using the regularization strength with the highest average accuracy from the cross-validation. This model was then evaluated on the test set. This procedure was repeated on 100 randomly generated training and test sets. Once model evaluation was completed, a separate ridge regression classifier was fit using a cross-validated search for an optimal regularization strength and all unlabeled audio recordings were classified. All interactions labelled as Male by this classifier were then split by user intent using text matching patterns (regex) to match commonly used words.

    \begin{figure}
        \centering
        \includegraphics[width=\linewidth]{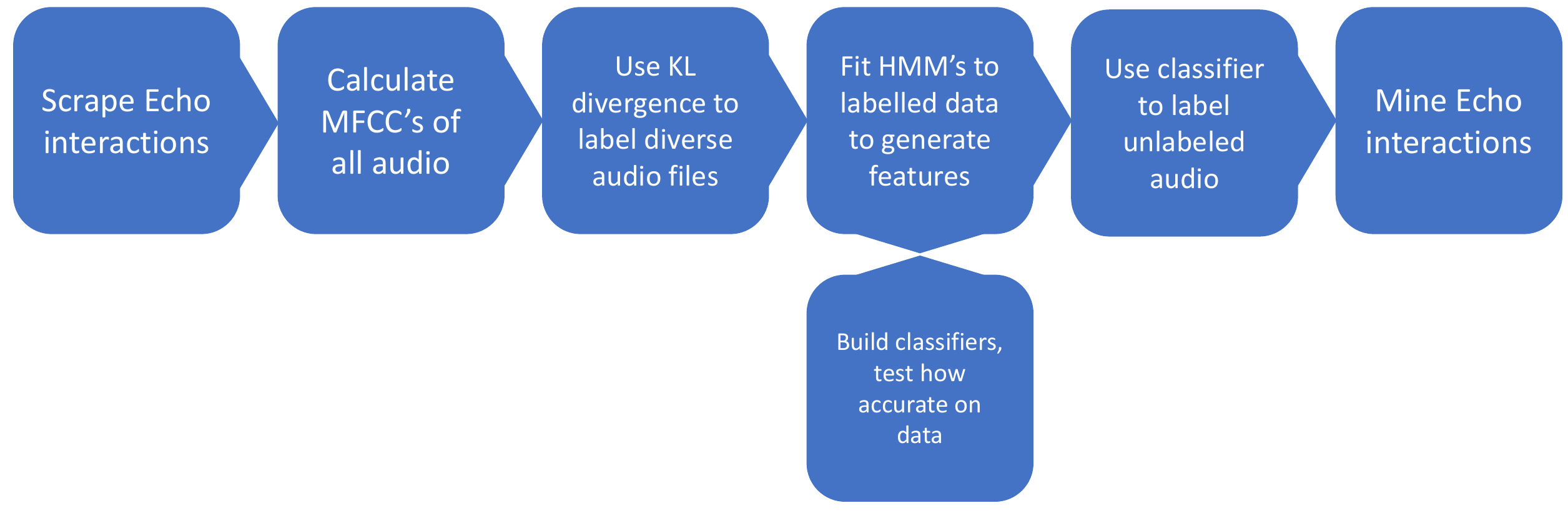}
            \caption{Counts of Amazon activity statuses for all interactions downloaded.}
            \label{fig:work_flow}
    \end{figure}
    
    \subsection{Machine Learning} 
    \label{sub:machine_learning}
\begin{figure}[htb]
        \centering
         \begin{subfigure}[b]{0.475\textwidth}
            \includegraphics[width=\linewidth]{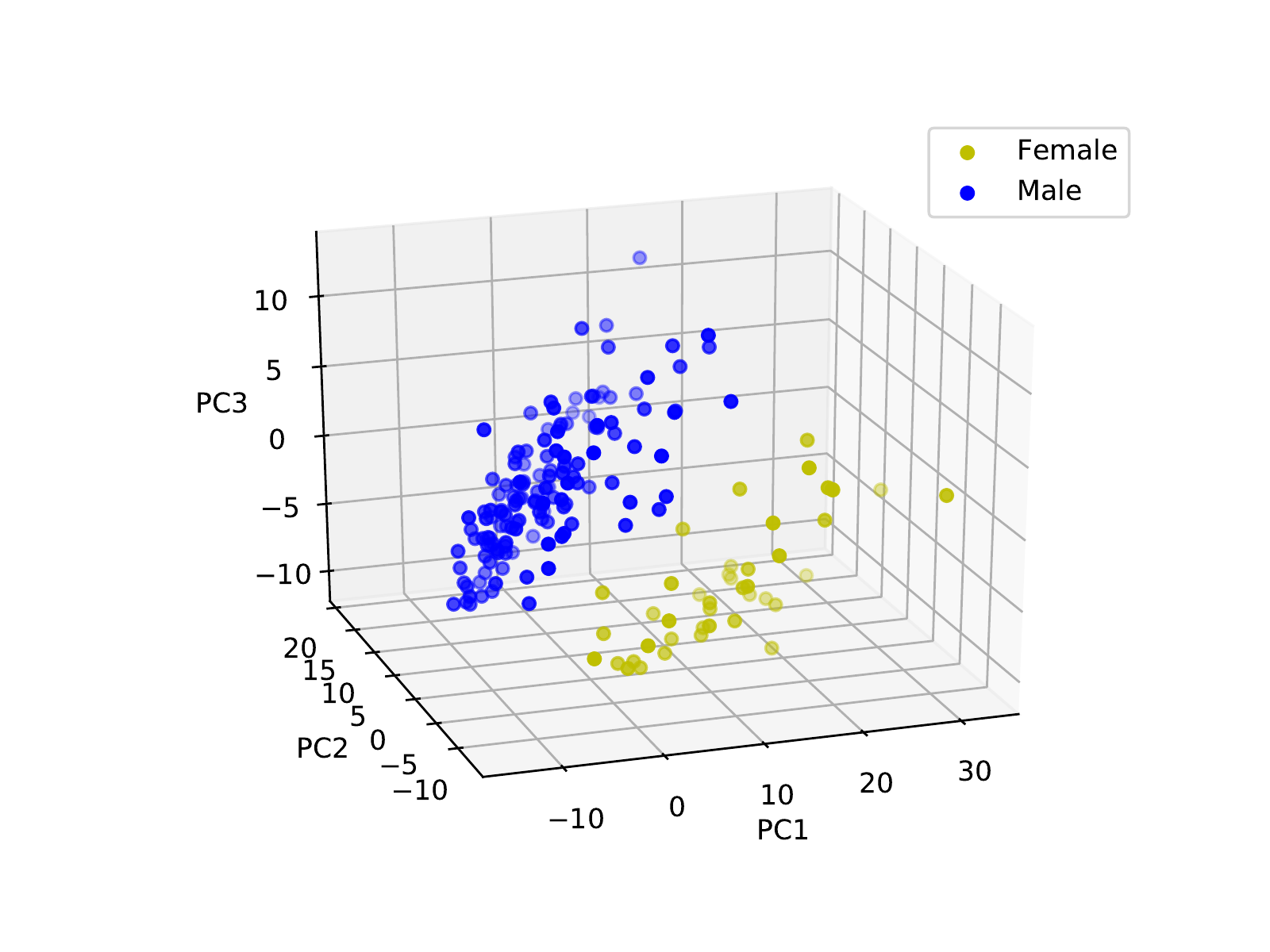}
            \caption{Scatter plot showing the first three principal components generated using all manually labeled data.}
            \label{fig:pca}
        \end{subfigure}
        \begin{subfigure}[b]{0.475\textwidth}
            \includegraphics[width=\linewidth]{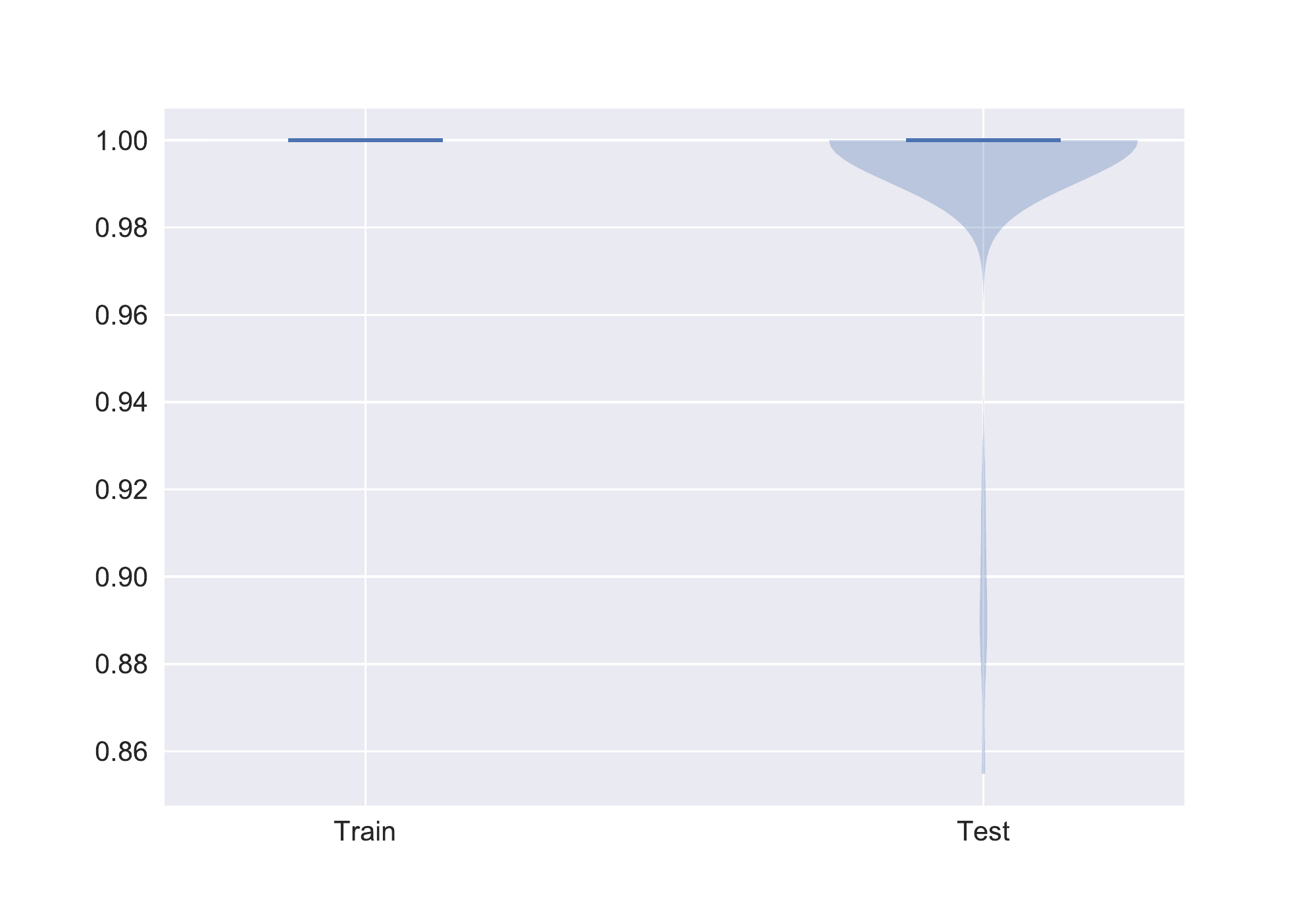}
            \caption{Violin plot of training and test accuracy accross 100 random training test splits. The median accuracies is show as a dark blue bar.}
            \label{fig:cross_validation}
         \end{subfigure}
         \caption{Machine learning results}
         \label{fig:machine_learning}
    \end{figure}

    A principal components analysis (\Cref{fig:pca}) of the similarity matrix extracted from Hidden Markov modeling showed that the two classes (Male and Female) are linearly separable, as the two classes are distinct in principal component space. Nested cross-validation with removal of features associated with the test set demonstrates high accuracy in both training and test sets (\Cref{fig:cross_validation}). All 100 folds performed with 100\% accuracy on the training set and only 21 folds have a test set accuracy of less than 100\%.    
    
    \subsection{Echo Usage} 
    Once model evaluation was completed, a separate ridge regression classifier was fit using a cross-validated search for an optimal regularization strength (alpha=100) and all unlabeled audio recordings were classified. All audio files classified as Male were then used to analyze how the Echoes were being used in the home. The most common task asked of the Echo was to listen to or control the playing of music (\Cref{tab:echo_usage}). The second most common task was the use of a timer, these timers were on average 13.4 minutes and most frequently (87.5\%) located in the kitchen so likely correspond to oven timers. A large amount (307) of the recorded audio was not transcribed successfully by the Alexa system (\Cref{tab:echo_usage}), these occur when the wake word was spoken and then nothing was said, the Echo determined that the wake word had been spoken but in fact it had not, or the Alexa system split the interaction into two interactions, one for the wake word and another for the user intent.

    \begin{table}[htb]
        \centering
        \caption{Echo usage counts and terms used to curate.}
        \label{tab:echo_usage}
        \begin{tabular}{lllp{35mm}} \thickhline
            Category       & Counts  & Regex                                          & Example                                   \\ \thickhline
            Timer          & 92      & timer                                          & set timer for five minutes                \\ 
            Volume control & 88      & volume                                         & volume eight                              \\
            Weather        & 80      & weather$|$rain$|$temperature                   & what's the weather like today             \\
            Music          & 351     & play$|$stop$|$pause$|$track$|$listen$|$skip    & play the smiths                           \\
            Shopping       & 69      & shopping listen                                & add milk to the shopping list             \\
            Calendar       & 7       & calendar                                       & what's on my calendar tomorrow            \\
            Sport          & 36      & football$|$score                               & what's the football score                 \\
            Error          & 307     & \^{}alexa$\$|$\^{}$\$$                         & N/A                                       \\
            Other          & 168     & N/A                                            & how much does a tablespoon of sugar weigh \\ \thickhline
        \end{tabular}
    \end{table}


    \section{Discussion} 
    Interactions collected from two Echoes in a household of two people have provided a large and rich dataset of transcribed audio from a naturalistic setting. The methods outlined will enable studies to derive quantitative data describing how people have used these devices, and provide transcribed audio which can be used to track subtle changes in memory or verbal functions. The web scraper described in this work is to the best of our knowledge the first method available to extract these data automatically, and it has been made freely available as part of ESCAPE under the MIT license \parencite{_ucl-mig/escape_????}. Participants' data is automatically collected and securely stored by Amazon, participants can then sign in to their own Amazon account and their data is then downloaded to a secure location. To curate Echo interactions into a useable dataset it was necessary to code which individual made each interaction, in this work we have used a machine learning approach. Using this approach, we have shown that a viable alternative to manually coding all data points, is to label a small but diverse dataset, train a classifier and propagate labels onto the rest of the dataset. Though these initial results show great promise for speaker recognition, a limitation in this work is the use of only one male and one female participant. This speaker recognition problem may be easier than problems involving participants of the same gender. After application of ESCAPE the resultant dataset was used to perform a simple analysis of usage in the home, showing the usage habits of the first author. We hope that this analysis highlights the potential use of ESCAPE to gather data from the home.

    \section{Conclusion} 
    \label{sub:outlook}

    IoT devices may represent an economical, mobile, and non-invasive method for improving independence and wellbeing for people with dementia. ESCAPE is currently being used to extract and analyze data being collected in a home-based evaluation of Amazon Echo utility with people with posterior cortical atrophy. It is likely that IoT devices will continue to grow in popularity for healthcare applications due to their potential mutual benefits to both people with dementia and researchers. With an increasing number and variety of devices comes an increasing amount of data that can be collected about people in their homes, providing an invaluable resource to track diseases. The methods described above make up the ESCAPE software to extract and perform speaker recognition on data collected by the Amazon Echo. ESCAPE is made freely available to download under the MIT license, meaning that persons are able to copy, modify, merge, publish, distribute, sublicense, and/or sell copies \parencite{_ucl-mig/escape_????}.

    \section*{Author Contributions} 
    \label{sec:author_contributions}
    NCF, EH, MPS, SC and DA were responsible for the conception of the work. NCF performed all analysis, wrote all the code made available and drafted the article. EH, MPS, SC and DA revised the article.

    \section*{Ethics}
    This study was carried out in accordance with the recommendations of the UCL Research Ethics Committee with written informed consent from all subjects

    \section*{Funding} 
    \label{sec:funding}
    \emph{Echoes Around the Home} is funded by a Social Science Plus grant from UCL’s Collaborative Social Science Domain. NCF, SC and DA are funded by EPSRC (EP/M006093/1). EH and MPS are funded by ESRC (ES/L001810/1). 

    \section*{Conflict of Interest Statement}
    The authors declare that the research was conducted in the absence of any commercial or financial relationships that could be construed as a potential conflict of interest.

    \section*{Acknowledgements}
    NCF would like to thank all participants of the \emph{Echoes Around the Home}, especially JP. NCF would also like to thank the POND team for helpful comments.

    \printbibliography

\end{document}